\documentclass[a4paper,12pt]{article}
\usepackage{a4}

\begin{document}

\Large
\begin{center}

   \vspace*{2ex}
     {\bf Supersymmetry and quantum games  } \\
             \vspace*{2ex}
              {\bf Jaroslav HRUBY}

   {\bf Institute of Physics AV CR,Czech Republic }
   {\bf e-mail: hruby.jar@centrum.cz  }

\vspace*{2ex} January 2007

\vspace*{2ex}
\begin{abstract}

     In this paper we show the connection between the supersymmetry and quantum games.

\end{abstract}

\end{center}

\vspace*{3ex}

\normalsize

\section{Quantum games}

Game theory is the study of rational decision making of competing
agents in some conflict situation. It was developed in the 1930's
in the context of economic theory and was first formalized by von
Neumann and Morgenstern in The Theory of Games and Economic
Behavior \cite{1}. Subsequently much work was done on game theory
and it is now a mature discipline used in areas such as the social
sciences, biology and engineering.

In recent time it is believed that quantum mechanics has the
potential to bring about a spectacular revolution in quantum game.

The combination of quantum mechanics and game theory promises to
improve the odds of winning and may provide new insight into novel
many areas as military warfare, anthropology, social psychology,
economics, politics, business, philosophy, biology, physics and
cryptology. With the recent interest in quantum computing and
quantum information theory, there has been an effort to recast
classical game theory using quantum probability amplitudes, and
hence study the effect of quantum superposition, interference and
entanglement on the agents' optimal strategies.

A natural question arise: can supersymmetric quantum mechanics
play a role in quantum game theory.

We know that a quantum game may be regarded as a family of games
provided by quantum correlations which are absent in classical
settings, and our geometric picture used to portray the
correlation-family in this paper turns out to be quite convenient,
especially for analyzing the phase structures of the game.

Here we want to show the connection of the supersymmetry and
quantum games, starting from the role of supersymmetric  quantum
mechanics for quantum information theory and then to concern on
the role of symmetries for quantum games.

The first ground for supersymmetric quantum mechanics was given in
superextension of nonlinear model in the year $1977$. Here the
supersymmetry Lagrangian was given in $(1+1)$ space-time
dimensions \cite{2} in~1977 and later \cite{3},\cite{4},\cite{5}:

\begin{equation}
 L =
\frac{1}{2}[(\partial_{n}\phi)^{2}-V^{2}(\phi)+
        \bar{\psi}(i+V'(\phi))\psi]    \label{v1}
\end{equation}

where $\phi$ was a solitonic Bose field and $\psi$ was a Fermi
field and $V(\phi)$ was some nonlinear potential.

Substituting into (\ref{v1}) the following restriction to $(1+1)$
space-time dimension

\begin{eqnarray}
  & \phi\rightarrow x(t) & \partial_{n}\rightarrow\partial_{t} \\
  & \bar{\psi}\rightarrow\psi^{\top}\sigma_{2} &
    i\rightarrow i\partial_{t}\sigma_{2}
\end{eqnarray}

where \(\psi={\psi_{1}\choose\psi_{2}}\) with components being
interpreted as anticommuting $c$-numbers, $\sigma_{h}$ denotes the
Pauli matrices, then from (\ref{v1}) the Lagrangian of the
supersymmetric quantum mechanics is obtained.

\begin{equation}
 L_{SSQM}=\frac{1}{2}[(\partial_{t}x)^{2}-V^{2}(x)+
          \psi^{\top}(i\partial_{t}+\sigma_{2}V'(x))\psi]
\label{v2}
\end{equation}

and the corresponding Hamiltonian has the known form

\begin{equation}
 H_{SSQM}=\frac{1}{2}p^{2}+\frac{1}{2}V^{2}(x)
   +\frac{1}{2}i[\psi_{1},\psi_{2}]V'(x)  \label{v3}
 \end{equation}

 which was later proposed as supersymmetric quantum mechanics
(SSQM) \cite{6}.

The role of supersymmetric qubit field theory as supersymmetric
extension of the Deutsch´s quantum field theory was presented in
the work \cite{7}.

Why we are interesting about the problems of symmetry in quantum
games is the role of entanglement for  quantum information theory
and quantum game theory.

In this introduction we give some compilation of evolution in
quantum game theory to show big potential for applications not
only in quantum information processing.

Short review of quantum game theory can be as follows:

Apart from unsolved problems in quantum information theory
\cite{8}, quantum game theory may be useful in studying quantum
communication since that can be considered as a game where the
objective is to maximize effective communication.Inspired by
quantum game theory, the new branch of economic mathematics,
econophysics, has attempted to model aspects of market places with
quantum games \cite{9},\cite{10}.

There has been recent work linking quantum games to the production
of algorithms for quantum computers \cite{11} that may lead to new
ways of approaching such algorithms. There is also a suggestion
that nature may be playing quantum games at the molecular level
\cite{12}.

The seminal work on quantum game theory by Meyer \cite{13} studied
a simple coin tossing game and showed how a player utilizing
quantum superposition could win with certainty against a classical
player. A general protocol for two player-two strategy quantum
games with entanglement was developed by Eisert et al \cite{14}
using the well known prisoners' dilemma as an example and this was
extended to multiplayer games by Benjamin and Hayden \cite{15}.
Since then more work has been done on quantum prisoners' dilemma
\cite{16},\cite{17},\cite{18},\cite{19},\cite{20},\cite{21},\cite{22}
and a number of other games have been converted to the quantum
realm including the battle of the sexes \cite{16, 23, 24}, the
Monty Hall problem \cite{21, 26, 27}, rock-scissors-paper
\cite{28}, and others \cite{15, 29, 30, 31, 32, 33, 34, 35, 36}.

In section two we introduce some terminology and basic ideas of
game theory. We will introduce the idea of quantum game theory and
give the general protocol for the construction of such games,
along with a review of the existing work in this field.

In sections three and four  we have been showing  the role of
supersymmetric quantum mechanics (SSQM)in quantum information
theory and the role of quantum Yang-Baxter equation and universal
quantum gate for entanglement.

In sections five we have been discussing symmetries in 2-player
games.

Finally, in conclusions we have been showing some questions and
suggestions.

\section{Terminology and basic ideas of game theory}

Game theory attempts to mathematically model a situation where
agents interact. The agents in the game are called players, their
possible actions moves, and a prescription that specifies the
particular move to be made in all possible game situations a
strategy. That is, a strategy represents a plan of action that
contains all the contingencies that can possibly arise within the
rules of the game. In response to some particular game situation,
a pure strategy consists of always playing a given move, while a
strategy that utilizes a randomizing device to select between
different moves is known as a mixed strategy. The utility to a
player of a game outcome is a numerical measure of the
desirability of that outcome for the player.

A payoff matrix gives numerical values to the players' utility for all the game outcomes.
The players are assumed to be rational and self-interested. That is, they will attempt to maximize their expected
utility by exploiting all the information available to them according to the rules of the game, but without recourse
to psychology or displays of altruism. Games in which the choices of the players are known as soon as they are made
are called games of perfect information. These are the main ones that are of interest to us here.

The definitions of utility and rationality are some what circular since, in practice, the method of establishing the
utility of a particular game outcome is by gauging the extent to which it is chosen by a rational player over other
outcomes. Does a rational player take into account that other players are also rational? If so, then the definition of
rationality becomes self-referential posing further problems for the concept.

A dominant strategy is one that does at least as well as any
competing strategy against any possible moves by the other
player(s). The Nash equilibrium (NE) is the most important of the
possible equilibria in game theory. It is the combination of
strategies from which no player can improve his/her payoff by a
unilateral change of strategy. The Pareto optimal outcome of the
game is the one which maximizes the sum of the players' payoffs.
An evolutionary stable strategy (ESS) [37] is one that, when
played by a population, is resistant against invasion by a small
group playing a "mutant" (i.e., a slightly altered) strategy. The
set of all strategies that are ESS is a subset of the NE of the
game. A two player, zero-sum game is one where the interests of
the players are diametrically opposed. That is, the sum of the
payoffs for any game result is zero. In such a game a saddle point
is an entry in the payoff matrix for (say) the row player that is
both the minimum of its row and the maximum of its column.

Some situations where there is only a single player along with chance elements can be treated by game theory as a
game against nature.

WE can present the famous example - the prisoners' dilemma (PD):

a two player game where each player has two possible moves is known as a 2x2 game, with obvious generalizations
to larger strategic spaces or number of players. As an example, consider one such game that has deservedly received
much attention: the prisoners' dilemma. Here the players' moves are known as cooperation (C) or defection (D). The
payoff matrix is such that there is a conflict between the NE and the Pareto optimal outcome.

The game is symmetric and there is a dominant strategy, that of always defecting, since it gives a better payoff if the other
player cooperates (five instead of three) or if the other player defects (one instead of zero).

Where both players have a dominant strategy this combination is
the NE. The NE outcome (D,D) is not such a good one for the
players, however, since if they had both cooperated they would
have both received a payoff of three, the Pareto optimal result.
In the absence of communication or negotiation we have a dilemma,
some form of which is responsible for much of the misery and
conflict through out the world.

Introductory ideas to quantum game theory can be demonstrated on a
penny flip One of the simplest gaming devices is that of a two
state system such as a coin. If we suppose we have a player than
can utilize quantum moves we can demonstrate how the expanded
space of possible strategies can be turned to advantage. Meyer, in
his seminal work on quantum game theory [13], considered the
simple game "penny flip" that consists of the following: Alice
prepares a coin in the heads state, Bob, without knowing the state
of the coin, can choose to either ip the coin or leave its state
unaltered, and Alice, without knowing Bob's action, can do
likewise. Finally, Bob has a second turn at the coin. The coin is
now examined and Bob wins if it shows heads. A classical coin
clearly gives both players an equal probability of success unless
they utilize knowledge of the other's psychological bias, and such
knowledge is beyond analysis by game theory.

To quantize this game, we replace the coin by a two state quantum system such as a spin one-half particle. Now
Bob is given the power to make quantum moves while Alice is restricted to classical ones. Can Bob profit from his
increased strategic space ?

Let $|0\rangle$
and $|1\rangle$ represent the "heads" state and  the "tails" state. Alice initially prepares the
system in the $|0\rangle$ state. Bob can proceed by first applying the Hadamard operator,

\begin{equation}
H = \frac{1}{\sqrt{2}} \left( \begin{array}{cc}
                              1 & 1 \\
                               1 & -1 \end{array} \right)
\label{v4}
\end{equation}

putting the system into the equal superposition of the two states:

\begin{equation}
\frac{1}{\sqrt{2}}{\left( |0\rangle  + |1\rangle \right) } ,
\label{v5}
\end{equation}

Now Alice can leave the "coin" alone or interchange the states $|0\rangle$
and $|1\rangle$, but if we suppose this is done without causing the system to decohere
either action will leave the system unaltered, a fact that can be exploited by Bob. In his second move he applies the
Hadamard operator again resulting in the pure state $|0\rangle$ thus winning the game. Bob utilized a superposition of states
and the increased latitude allowed him by the possibility of quantum operators to make Alice's strategy irrelevant,
giving him a certainty of winning.
We shall see later that quantum enhancement often exploits entangled states, but in this case it is just the increased
possibilities allowed the quantum player that proved decisive. Du et al has also considered quantum strategies in a
simplified card game that do not rely on entanglement [29].

It can be generealized to a  general prescription:
where a player has a choice of two moves they can be encoded by a single bit. To translate this into the quantum
realm we replace the bit by a quantum bit or qubit that can be in a linear superposition of the two states. The basis
states $|0\rangle$ and $|1\rangle$ correspond to the classical moves. The players' qubits are initially prepared in some state to be
specified later. We suppose that the players have a set of instruments that can manipulate their qubit to apply their
strategy without causing decoherence of the quantum state. That is, a pure quantum strategy is a unitary operator
acting on the player's qubit. After all players have executed their moves a measurement in the computational basis
({$|0\rangle$,$|1\rangle$}) is made on the final state and the payoffs are determined in accordance with the payoff matrix. Knowing the
final state prior to the measurement, the expectation values of the payoffs can be calculated. The identity operator
I corresponds to retaining the initial choice while

\begin{equation}
F=i{\sigma}_1= \left( \begin{array}{cc}
                              0& i \\
                               i& 0 \end{array} \right),
\label{v6}
\end{equation}

corresponds to a bit flip F. The resulting quantum game should contain the classical one as a subset.
We can extend the list of possible quantum actions to include any physically realizable action on a player's qubit that
is permitted by quantum mechanics. Some of the actions that have been considered include projective measurement
and entanglement with ancillary bits or qubits.

A quantum game of the above form is easily realized as a quantum algorithm. Physical simulation of such an
algorithm has already been performed for a quantum prisoners' dilemma in a two qubit nuclear magnetic resonance
computer [18].

How it can be applied in cryptology ?  If we suppose that
classical cryptanalysis with brute force attack where the player
(codebreaker) is trying all possible keys. Every unsuccessful
application of a key (for example to decipher a symmetric cipher)
represents a lose for the player. The coherence, entanglement and
superposition in the quantum case give chance for the winning
strategy in the decoding game. In the quantum case can also the
coherent quantum effects impede the players to such an extent that
the classical game outperforms the quantum game.

  A particularly interesting application may arise from an effect, so called Parondo effect,
let it takes another look at the PD game . PD game can be resolved
if the players resort to strategies available in quantum theory.
Randomly switching between two prisoners' dilemma games in which
the players "lose" should prompt the the players to choose Pareto
optimal and thus increase their global gain (i.e.they now win).

The combination this strategy with the Grover search algorithm can
give a new way to attack on symmetric cipher. The research in this
direction is in progress.

 Subsequently, Marinatto and Weber
\cite{44} examined the dilemma in the Battle of the Sexes (BoS)
game, another typical dilemma in game theory,  and observed that
this, too, could be resolved by adopting a quantum strategy
involving a maximally entangled state. Application of quantum
strategies to various other games, such as the Stag Hunt (SH) or
the Samaritan's Dilemma game, has also been discussed in
\cite{45}.

\section{SSQM and quantum information}

Unitary gates, which play crucial role in QIS, are connected with
R-matrices from quantum Yang-Baxter equation (QYBE)\cite{46} and
via Yang-Baxterization Hamiltonians can be expressed as square
root of b-matrices. This established the connection between
topology and QM, especially it gives topological interpretation of
entanglement. Here we show it on the two-qubit state.

To understand this we start with a simple idea of quantum
algorithm square root of not
 via partner--superpartner case, which appears in supersymmetric
 quantum mechanics (SSQM).
Supersymmetry is special case of anyonic Lie algebra
$C[\theta]/{\theta}^n$ with one coordinate $\theta ,
{\theta}^n=0$(for our Grassmann variable ${\Theta}^2 = 0$) with
anyonic variables fulfil:
\begin{equation}
\theta''=\theta +\theta'  ,  \theta'\theta = e^{\frac{2\pi
i}{n}}\theta\theta'  \label{v7}
\end{equation}
 and the category of n-anyonic vector spaces where objects are
 $Z_n$-graded spaces with the braided transposition:

\begin{equation}
\Psi(x \otimes y)= e^{{\frac{2\pi i}{n}\mid x\parallel y\mid}}y
\otimes x  \label{v8}
\end{equation}
 on elements $x,y$ of homogenous degree $\parallel$.

 Thus an anyonic braided group means as $Z_/n$-graded algebra B
 and coalgebra defined as $\varepsilon$:
 \begin{equation}
 B\rightarrow C , \triangle(x)=x_A\otimes x^A ,\label{v9}
\end{equation}
where summation understood over terms labelled by  A of
homogeneous degree.

Coassociative and counital is in the sense:
 \begin{equation}
 x_{AB}\otimes {x_A}^B \otimes x^A = x^A\otimes {X^A}_B\otimes x^{AB} ,
 \varepsilon(x_A)x^A =x = x_A\varepsilon (x^A) \label{v10}
\end{equation}
and obeying
 \begin{equation}
 (xy)_A\otimes (xy)^A = x_Ax_B\otimes x^Ax^Be^{{\frac{2\pi i}{n}\mid x^A\parallel
 x_B\mid}}, \label{4.22}
\end{equation}
for all ${x,y}\in B$.The axioms for the antipode are as for usual
quantum groups \cite{6}. There is shown that anyonic calculus and
anyonic matrices are generalization of quantum matrices and
supermatrices.

The $N^2$ generators ${t^i}_j= f(i)-f(j)$, where f is a degree
$Z_/n$ associated with the row or column fulfil
\begin{equation}
e^{(\frac{2\pi i}{n})\{f(i)f(k)+f(j)f(l)\}}{{{\Re^i}_a}^k}_b
{t^a}_b {t^b}_l=e^{(\frac{2\pi i}{n})\{f(j)f(l)+f(i)f(k)\}}{t^k}_b
{t^i}_a{{{\Re^a}_j}^b}_l, \label{4.23}
\end{equation}

\begin{equation}
\triangle {t^i}_j = {t^i}_a\otimes {t^a}_j, \label{4.24}
\end{equation}

\begin{equation}
\varepsilon {t^i}_j = {{\delta}^i}_j \label{4.25}
\end{equation}

 It is required that $\Re$ obeys certain anyonic QYBE.
 One method to obtain $\Re$ is start with certain unitary solution
 R of the usual QYBE and "transmute: them.
 In the diagrammatic notation this braided mathematics is known \cite
 {46}. In QIS a "wiring" notation (which is known in physics like Feynman
 diagrams) was used for information flows for example for
 teleportation.
 In the diagrammatic notation we "wire" the outputs of maps into
 the inputs of other maps to construct the algebraic operation
Information flows along these wires except that under and over
crossing are nontrivial operators , say $U$ and $U^{-1}$. Such
operators can be universal quantum gates in QIS. Generally in
anyonic we have new richer kind of "braided quantum field
information mathematics".

To show it we begin  with coincidence of the supersymmetric square
root in SSQM $n=2$ and square root of the not gate in QIS as an
illustrative example.

It is well known SSQM is generated by supercharge operators
$Q^{+}$ and $Q^{-}=(Q^{+})^{+}$ which together with the
Hamiltonian $H=2H_{SSQM}$ of the system, where

\begin{eqnarray*}
 H_{SSQM} & = & \frac{1}{L}\left(
     \begin{array}{cc}
       -\frac{d^{2}}{d\,t^{2}}+v^{2}+v' & 0 \\
       0 & -\frac{d^{2}}{d\,t^{2}}+v^{2}-v'
     \end{array}  \right)    \\
 H & = & \left(
     \begin{array}{cc}
       H_{0} & 0 \\ 0 & H_{1}
     \end{array}  \right) =
   \left(
     \begin{array}{cc}
       A^{+}A^{-} & 0 \\ 0 & A^{-}A^{+}
     \end{array}  \right) =
     -\left( \frac{d^{2}}{dx^{2}}\right)+\sigma_{3}v'
\end{eqnarray*}

fulfil the superalgebra

\begin{equation}
{Q^{\pm}}^{2}=0 \ , \ \ [H,Q^{-}]=[H,Q^{+}] \ ,
                              \ \ H=\{Q^{+},Q^{-}\}=Q^{2}\, \label{4.26}
\end{equation}

where

\[ Q^{-}=\left(
     \begin{array}{cc}
       0 & 0 \\  A^{-} & 0
     \end{array}  \right) \ , \
 Q^{+}=\left(
     \begin{array}{cc}
       0 & A^{+} \\  0 & 0
     \end{array}  \right)   \ , \  \  Q = Q^{+}+Q^{-} \]
and

\begin{equation}
 A^{\pm}=\pm\frac{d}{dx}+v(x) \ , \ \ v'=\frac{dv}{dx}  . \label{4.27}
\end{equation}

Such Hamiltonians \(H_{0},H_{1}\) fulfil
\begin{equation}
H_{0}A^{+}=A^{+}H_{1} \ , \ \ A^{-}H_{1}=H_{0}A^{-}.\label{4.28}
\end{equation}
The eigenfunctions of the Hamiltonian
\begin{equation}
H  =  \left(     \begin{array}{cc}
       H_{0} & 0 \\ 0 & H_{1}
     \end{array}  \right)
\end{equation}

 are  $\phi=(|0\rangle,|1\rangle)^T$ and then

\begin{equation}
     Q^0\phi=|1\rangle,
\end{equation}

\begin{equation}
     Q^1\phi=|0\rangle.
\end{equation}

 where we denote $Q^-=Q^0$and$Q^+=Q^1$ ,respectively.

The previous relations in lead to the double degeneracy of all
positive energy levels, of belonging to the ``0'' or ``1'' sectors
specified by the grading state operator $S=\sigma_{3}$, where
\[[S,H]=0 \ \mbox{and}\ \{S,Q\}=0 \ . \]

The $Q$ operator transforms eigenstates with \(S=+1\), i. e. the
null-state $|0\rangle$  into eigenstates with \(S=-1\), i. e. the
one-state $|1\rangle$ and vice versa.

With this notation the square root of not gate $M_- = \sqrt M_-
\sqrt M_-$ is represented by the unitary matrix $\sqrt M_-$:
\[\sqrt M_- =\frac{1}{2}\left[
     \begin{array}{cc}
       1+i & 1-i \\  1-i & 1+i
     \end{array}  \right] \ , \]
that solves:
\begin{equation}
  \sqrt M_- \sqrt M_- |0\rangle =\sqrt M_- (\frac{1+i}{2}|0\rangle + \frac{1-i}{2} |0\rangle)  = |1\rangle ,
    \end{equation}
\begin{equation}
  \sqrt M_- \sqrt M_- |1\rangle  =  |0\rangle
\end{equation}

In such way this supersymmetric double degeneracy represents two
level quantum system and we can see the following:

\begin{eqnarray*}
 Q & = & Q^{+}+Q^{-}= \left(
     \begin{array}{cc}
       0 & A^{+} \\  A^{-} & 0
     \end{array}  \right) \ , \\
  \tau & = & \sigma_{3} = \left(
     \begin{array}{cc}
       1 & 0\\  0 & -1
     \end{array}  \right) \ , \\
 \{Q,\tau\} & = & 0 \ .
\end{eqnarray*}

It implies that the operator supercharge $Q$ really transforms the
state \(|0\rangle,\ |1\rangle\) as the operator square root of not
quantum algorithm operator $M_-$. In such a way supersymmetric
``square root'' corresponds the ``square root of not'' in QIS. We
can ask if generally the Hamiltonians of the unitary braiding
operators which leads to the Schr\"{o}dinger equations are square
root of QIS unitary gates. Such QIS unitary gates are unitary
solutions of QYBE. The answer is positive and it can be explicitly
shown for two-qubit system.

   A system of two quantum bits is four dimensional space $H_4 = H_2 \otimes H_2$
having orthonormal basis $|00\rangle, |01\rangle, |10\rangle,
|11\rangle $ .

    A state of two-qubit system is a unit-length vector
		
\begin{equation}
  {a}_0 |00\rangle  + {a}_1 |10\rangle + {a}_2 |01\rangle  + {a}_3 |11\rangle  ,
\end{equation}

so it is required ${|{a}_0|}^2 + {|{a}_1|}^2 + {|{a}_2|}^2 +
{|{a}_3|}^2  = 1$.

We can see a state $z\in H_4$

\begin{equation}
  z=\frac{1}{2}( |00\rangle  + |01\rangle +  |10\rangle  + |11\rangle )= \frac{1}{\sqrt{2}}(|0\rangle + |1\rangle)\frac{1}{\sqrt{2}}(|0\rangle + |1\rangle) ,
\end{equation}

of a two-qubit system is decomposable, because it can be written
as a product of states in $H_2$. A state that is not decomposable
is entangled.

 Consider the unitary matrix

 \begin{equation}
 \overline{R}= \left( \begin{array}{cccc}
                              a_0 & 0 & 0 & 0\\
                              0 & 0 & a_3 & 0\\
                              0 & a_2 & 0 & 0\\
                              0 & 0   & 0 & a_1   \end{array} \right) ,
\end{equation}

defines a unitary mapping, whose action on the two-qubit basis is

\begin{equation}
    \Psi=\overline{R}(\psi\otimes\psi) =       \overline{R} \left( \begin{array}{c}
                              |00\rangle\\
                              |01\rangle\\
                              |10\rangle\\
                              |11\rangle  \end{array}\right) = \left( \begin{array}{c}
                              a_0|00\rangle\\
                              a_3|10\rangle\\
                              a_2|01\rangle\\
                              a_1|11\rangle  \end{array}\right) . \label{1.2}
\end{equation}

For $a_0a_1 \neq a_2a_3$ the state is entangled.For example the
state is entangled $\frac{1}{\sqrt{2}}(|10\rangle +|01\rangle)$ is
entangled.

\section{The QYBE and universal quantum gate }

    Matrix

\begin{equation}
 M_{CNOT}= \left( \begin{array}{cccc}
                              1 & 0 & 0 & 0\\
                              0 & 1 & 0 & 0\\
                              0 & 0 & 0 & 1\\
                              0 & 0 & 1 & 0  \end{array} \right) ,
\end{equation}

defines a unitary mapping, whose action on the two-qubit basis is

\begin{equation}
 M_{CNOT} \left( \begin{array}{c}
                              |00\rangle\\
                              |01\rangle\\
                              |10\rangle\\
                              |11\rangle  \end{array}\right) = \left( \begin{array}{c}
                              |00\rangle\\
                              |01\rangle\\
                              |11\rangle\\
                              |10\rangle  \end{array}\right) . \label{1.2}
\end{equation}

    Gate $M_{CNOT}$ is called controlled not, since the target qubit
    is flipped if and only if the control bit is 1 and the gate
    $M_{CNOT}$ plays important role for entanglement.

 Let R
\begin{equation}
 R   = \frac{1}{\sqrt{2}} \left( \begin{array}{cccc}
                              1 & 0 & 0 & 1\\
                              0 & 1 &-1 & 0\\
                              0 & 1 & 1 & 0\\
                              1 & 0 & 0 & 1  \end{array} \right) ,
\label{1.6}
\end{equation}
be the unitary solution to the QYBE.

Let $M = M_1\bigotimes M_2$, where

\begin{equation}
 M_1   =  \frac{1}{\sqrt{2}}\left( \begin{array}{cc}
                              1 & 1\\
                              1 &-1  \end{array} \right) ,
\end{equation}

and

\begin{equation}
 M_2   =  \frac{1}{\sqrt{2}}\left( \begin{array}{cc}
                              -1 & 1\\
                               i & i  \end{array} \right) ,
\label{1.7}
\end{equation}

Let $N = N_1\bigotimes N_2$, where

\begin{equation}
 N_1   =  \frac{1}{\sqrt{2}}\left( \begin{array}{cc}
                              1 & i\\
                              1 &-i  \end{array} \right) ,
\end{equation}

and

\begin{equation}
 N_2   = - \frac{1}{\sqrt{2}}\left( \begin{array}{cc}
                              1 & 0\\
                              0 & i  \end{array} \right) ,
\label{1.7}
\end{equation}

Then $M_{CNOT} = M\cdot R\cdot N$ can be expressed in terms of R.
 In the QYBE solution R-matrices usually depends on the
 deformation parameter q and the spectral parameter x. Taking the
 limit of $x\longrightarrow0$ leads to the braided relation from
 the QYBE and the BGR b-matrix from the R-matrix. Yang-
 Baxterization is construct the $R(x)$ matrix from a given BGR
 b-matrix.
    The BGR b of the eight-vertex model assumes the form

\begin{equation}
 b\pm   = \left( \begin{array}{cccc}
                              1 & 0 & 0 & q\\
                              0 & 1 & \pm1 & 0\\
                              0 & \mp1 & 1 & 1\\
                        -q^{-1} & 0    & 0 & 1  \end{array} \right) ,
\end{equation}

    It has two eigenvalues $\Lambda_{1,2}= 1\pm i $. The
    corresponding $R(x)$-matrix via Yang-Baxterization is obtained
    to be
		
\begin{equation}
 R_{\pm}(x) = b + x \Lambda_1 \Lambda_2 =
 \left( \begin{array}{cccc}
                              1+x & 0 & 0 & q(1-x)\\
                              0 & 1+x & \pm(1-x) & 0\\
                              0 & \mp(1-x) & 1+x & 1\\
                    -q^{-1}(1-x)& 0    & 0 & 1+x  \end{array} \right),
\end{equation}

    Introducing the new variables $\theta,\varphi$ as follows
		
\begin{equation}
\cos\theta = \frac{1}{\sqrt{1+x^2}}, \sin \theta =
\frac{x}{\sqrt{1+x^2}}, q = e^{i\varphi}
\end{equation}

the R-matrix has the form

\begin{equation}
R_\pm (\theta) = \theta\cos(\theta)b_{\pm}(\varphi) +
\sin(\theta)b_{\pm}^{-1}(\varphi)
\end{equation}

The time-independent Hamiltonian $H_\pm$ has the form

\begin{equation}
 H_\pm = -\frac{i}{2}{b_\pm}^2(\varphi) = \frac{i}{2} \left( \begin{array}{cccc}
                              0 & 0 & 0 & -e^{i\varphi}\\
                              0 & 0 & \mp1 & 0\\
                              0 & \pm1 & 0 & 0\\
                        e^{-i\phi} & 0    & 0 & 0  \end{array} \right) ,
\end{equation}

It is fundamental to view unitary braiding operators describing
topological entanglements as universal quantum gates for quantum
computation.

Entanglement is playing main role in quantum game theory and via
braiding is connected with supersymmetry in the following sense:

a generalization of the category of super vector spaces in which
the grading is replaced by a braiding is the braided category and
in this sense one way to study topological entanglement and
quantum entanglement is to try making direct correspondences
between patterns of topological linking and entangled quantum
states.

It is well known that the $2\otimes2$-dimensional quantum systems
are of particular interest and importance to the study of quantum
games.

In particular, such systems are considered as the natural
requirement for playing two-player quantum games. Eisert et. al
used a $2\otimes2$ system to investigate the impact of
entanglement on a NE in Prisoners' Dilemma.

Let $\mathcal{H}_{A}$ and $\mathcal{H}_{B}$ be two-dimensional
Hilbert spaces with bases $\left\{  \left|  0\right\rangle
_{A},\left|  1\right\rangle _{A}\right\}  $ and$\ \left\{  \left|
0\right\rangle _{B},\left| 1\right\rangle _{B}\right\}  $,
respectively. Then a basis for the Hilbert space
$\mathcal{H}_{A}\otimes\mathcal{H}_{B}$ is given by

\begin{equation}
\left\{  \left|  0\right\rangle _{A}\otimes\left|  0\right\rangle
_{B},\ \left| 0\right\rangle _{A}\otimes\left|1\right\rangle _{B},
\ \left|1\right\rangle _{A}\otimes\left| 0\right\rangle _{B},\
\left| 1\right\rangle _{A}\otimes\left| 1\right\rangle
_{B}\right\}  .
\end{equation}

The most general state in the Hilbert space
$\mathcal{H}_{A}\otimes \mathcal{H}_{B}$ can be written as

\begin{equation}
\left|  \psi\right\rangle _{AB}=\sum_{i,j=0}^{1}c_{ij}\left|
i\right\rangle _{A}\otimes\left|  j\right\rangle _{B}
\end{equation}

which is usually written as $\left|  \psi\right\rangle =\sum_{i,j}%
c_{ij}\left|  ij\right\rangle $. Here the indices $i$ and $j$
refer to states in the Hilbert spaces $\mathcal{H}_{A}$ and
$\mathcal{H}_{B}$, respectively. The general normalized states in
$\mathcal{H}_{A}$ and $\mathcal{H}_{B}$ are $\left|
\psi^{(A)}\right\rangle _{A}=c_{0}^{(A)}\left|  0\right\rangle
+c_{1}^{(A)}\left|  1\right\rangle $ and $\left|
\psi^{(B)}\right\rangle _{B}=c_{0}^{(B)}\left|  0\right\rangle
+c_{1}^{(B)}\left|  1\right\rangle $,
respectively. The state $\left|  \psi\right\rangle _{AB}$ is a product state when%

\begin{equation}
\left|  \psi\right\rangle _{AB}=(c_{0}^{(A)}\left|  0\right\rangle
+c_{1}^{(A)}\left|  1\right\rangle )\otimes(c_{0}^{(B)}\left|
0\right\rangle +c_{1}^{(B)}\left|  1\right\rangle )
\label{Separability Criterion for 2-qubit state}%
\end{equation}

where $\left|  c_{0}^{(A)}\right|  ^{2}+\left|  c_{1}^{(A)}\right|
^{2}=1$ and $\left|  c_{0}^{(B)}\right|  ^{2}+\left|
c_{1}^{(B)}\right|  ^{2}=1$.

For example, consider the state%

\begin{equation}
\left|  \psi\right\rangle _{AB}=(\left|  00\right\rangle +\left|
11\right\rangle )/\sqrt{2} \label{2-qubit entangled state}%
\end{equation}

For this state the separability criterion for 2-qubit states
implies the results%

\begin{equation}
c_{0}^{(A)}c_{0}^{(B)}=1/\sqrt{2},\ \ c_{1}^{(A)}c_{1}^{(B)}%
=1/\sqrt{2},\ \ c_{0}^{(A)}c_{1}^{(B)}=0,\ \ c_{1}^{(A)}%
c_{0}^{(B)}=0
\end{equation}

These equations cannot be true simultaneously. The state is,
therefore, entangled.

\section{Symmetries in 2-player games }

To begin with, we first recapitulate the classical 2-player,
2-strategy game and then introduce its quantum version following
\cite{48}. Let $i = 0, 1$, $j = 0, 1$ be the labels of the
strategies available for the players, Alice and Bob, respectively,
and let also $A_{ij}$ and  $B_{ij}$ be their payoffs when their
joint strategy is $(i, j)$. In classical game theory, the game is
said to be \lq symmetric\rq\ if $B_{ji} = A_{ij}$, that is,  if
their payoffs coincide when their strategies are swapped $(i, j)
\to (j, i)$.

To make a distinction from the other symmetry discussed shortly,
we call such a game {\it $S$-symmetric} in this paper.

Similarly, we call the game {\it $T$-symmetric} if $B_{1-j, 1-i} =
A_{ij}$, that is, if the payoff matrices coincide when the
strategies of the two players are \lq twisted\rq\ as $(i, j) \to
(1-j, 1-i)$.

The BoS is an example of $T$-symmetric games with the additional
property $A_{01} = A_{10}$. The payoffs in these $S$-symmetric and
$T$-symmetric games are displayed in the form of the bi-matrix
$(A_{ij}, B_{ij})$ in Table .

Given a payoff matrix, each player tries to maximize his/her
payoff by choosing the best possible strategy, and if there exists
a pair of strategies in which no player can bring him/her in a
better position by deviating from it unilaterally, we call it a NE
of the game.   The players will be happy if the NE is unique and
fulfills certain conditions attached to the game ({\it e.g.},
Pareto-optimality or risk-dominance as mentioned later). Even when
there are more than one NE, the players  will still be satisfied
if a particular NE can be selected over the other upon using some
reasonings.  Otherwise, the players may face a dilemma, as they do
in the case of the in the Battle of the Sexes (BoS) game BoS and
the PD.

To introduce a quantum version of the classical game, we first
regard Alice's strategies $i$ as vectors in a Hilbert space ${\cal
H}_A$ of a qubit.  Namely, corresponding to the classical
strategies $i$  we consider vectors $\left| i \right>_A$ for $i =
0$ and $1$ which are orthonormal in ${\cal H}_A$.   A general
quantum strategy available for Alice is then represented by a
normalized vector $\left| \alpha \right>_A$ (with the overall
phase ignored, {\it i.e.}, a unit ray) in ${\cal H}_A$. Bob's
strategy is similarly represented by a normalized vector $\left|
\beta \right>_B$ in another qubit Hilbert space ${\cal H}_B$
spanned by orthonormal vectors $\left| j \right>_B$ for $j = 0$
and $1$ in ${\cal H}_B$. The strategies of the players can thus be
expressed in the linear combinations,

\begin{eqnarray*}
\begin{array}{cc}
\label{basis}
\left | \alpha \right>_A &=  \sum_i \xi_i(\alpha) \left | i \right>_A , \\
\left | \beta \right>_B &=     \sum_j \chi_j(\beta) \left | j
\right>_B ,
\end{array}
\end{eqnarray*}

using the bases $\left| i \right>_A$,  $\left| j \right>_B$
which correspond to the classical strategies, with complex
coefficients $\xi_i(\alpha)$, $\chi_j(\beta)$ which are functions
of the parameters $\alpha$ and $\beta$ normalized as $\sum_i \vert
\xi_i \vert^2 =  \sum_j \vert \chi_j \vert^2 = 1$.

The strategies of the individual players are, therefore, realized
by local actions implemented by the players independently.

\begin{table}[t]
\begin{center}
$S-symmetric$: \setlength{\doublerulesep}{0.1pt}
\begin{tabular}{ c|c c }
\hline\hline
strategy & Bob 0 & Bob 1 \\
\hline
Alice 0 & $\,(A_{00} , A_{00})$ &$( A_{01} , A_{10})$ \\
Alice 1 & $\,(A_{10} ,A_{01})$ &$(A_{11} ,A_{11})$ \\
\hline\hline
\end{tabular}
\end{center}

\end{table}
\begin{table}[t]
\begin{center}
$T-symmetric$: \setlength{\doublerulesep}{0.1pt}
\begin{tabular}{c|cc}
\hline\hline
strategy & Bob 0 & Bob 1 \\
\hline
Alice 0 & $\,(A_{00} , A_{11})$ &$( A_{01} , A_{01})$\\
Alice 1&  $\,(A_{10} ,A_{10})$ & $(A_{11} ,A_{00})$\\
\hline\hline
\end{tabular}
\end{center}
\caption{Payoff bi-matrices $(A_{ij}, B_{ij})$ of the
$S$-symmetric game (above) and the $T$-symmetric game (below).}
\end{table}

The {\it joint} strategy of the players, on the other hand,  is
given by a  vector in the direct product Hilbert space
$\mathcal{H} = \mathcal{H}_A \otimes \mathcal{H}_B$.

Here lies one of the crucial differences between the classical and
quantum games: in quantum game theory, the joint strategy is
specified not just by the choice of the strategies of the players
but also by furnishing the quantum correlation (essentially the
entanglement) between the individual strategies. Consequently, the
outcome of a quantum game rests also on a third party (or referee)
that determines the correlation.   To be more explicit, using the
product vector $| \alpha, \beta  \rangle = |  \alpha \rangle_A |
\beta \rangle_B$ which is uniquely specified by the individual
strategies, a vector in the total strategy space $\mathcal{H}$ is
written as $ \left | \alpha, \beta; \gamma \right> = J(\gamma)
\left | \alpha, \beta \right> = J(\gamma) \left | \alpha \right>_A
\left | \beta \right>_B, \label{qstate}$ where $J(\gamma)$ is a
unitary operator providing the quantum correlation between the
individual strategies.

The  correlation factor $J(\gamma)$ with the parameter set
$\gamma$ is designed to exhaust all possible joint strategies
available in ${\cal H}$. The payoffs for Alice and Bob are then
given by the expectation values of some appropriate self-adjoint
operators $A$ and $B$, respectively:

\begin{eqnarray*}
\begin{array}{cc}
\Pi_A(\alpha, \beta; \gamma) =   \left < \alpha, \beta; \gamma  |
A  |  \alpha,\beta; \gamma  \right > , \Pi_B(\alpha, \beta;
\gamma)
 =   \left < \alpha, \beta; \gamma  | B  |  \alpha, \beta; \gamma  \right > .
\label{payoff}
\end{array}
\end{eqnarray*}

To sum up, a quantum game is defined formally by the triplet
$\{{\cal H}, A, B\}$.

To choose the payoff operators $A$ and $B$, we require that, in
the absence of quantum correlations $J(\gamma) = I$ ($I$ is the
identity operator in ${\cal H}$), the payoff values reduce to the
classical ones when the players choose the \lq semiclassical
(pure) strategies\rq\ $ \left| i, j \right> = |  i \rangle_A | j
\rangle_B$,
\begin{equation}
\left< i', j' \right| A  \left| i, j \right> =  A_{ij} \delta_{i'
i}\delta_{j' j}, \left< i', j' \right| B  \left| i, j \right> =
B_{ij} \delta_{i' i}\delta_{j' j}. \label{classicalpayoff}
\end{equation}

Adopting, for simplicity,  the value $\gamma = 0$ for the
reference point  at which $J(\gamma) = I$ holds, we find that, for
the uncorrelated product strategies $|  \alpha, \beta; 0  \rangle
= | \alpha, \beta  \rangle$, the payoffs (\ref{payoff}) become
\begin{eqnarray*}
\begin{array}{cc}
&\Pi_A(\alpha, \beta; 0) = \left < \alpha, \beta; 0  | A  |
\alpha, \beta; 0  \right >  = \sum_{i,j} x_i  A_{ij} y_j,
\\
&\Pi_B(\alpha, \beta; 0) = \left < \alpha, \beta; 0  | B  |
\alpha, \beta; 0 \right >  = \sum_{i,j} x_i  B_{ij} y_j,
\label{clpayoff}
\end{array}
\end{eqnarray*}
 where $x_i = \vert \xi_i\vert^2$, $y_j = \vert \chi_j\vert^2$
represent the probability of realizing the strategies $ \left | i
\right>_A$, $ \left | j \right>_B$ under the general choice $
\left | \alpha \right>_A$, $ \left | \beta \right>_B$ .

This ensures the existence of a {\it classical limit} at which the
quantum game reduces to the classical game defined by the payoff
matrix $A_{ij}$, where now Alice and Bob are allowed to adopt {\it
mixed strategies}  with probability distributions $x_i$, $y_j$
($\sum x_i = \sum y_j = 1$) for strategies $i$, $j$.

We thus see that the quantum game is an extension of the classical
game, in which the correlation parameter $\gamma$ plays a role
similar to the Planck constant $\hbar$ in quantum physics in the
technical sense that the classical limit is obtained by their
vanishing limit.

Note that, since  the set $\{  \left| i, j \right>, \,  i, j = 0,
1\}$ forms a basis set in the entire Hilbert space ${\cal H}$, the
payoff operators $A$ and $B$ are uniquely determined from the
classical payoff matrices by; in other words, our quantization is
{\it unique}.

The aforementioned symmetries in classical game can also be
incorporated into quantum game by using corresponding appropriate
symmetry operators.  Indeed, by introducing the swap operator $ S
| i, j  \rangle = | j, i  \rangle, $ we see immediately that in
the classical limit the game is $S$-symmetric, $\Pi_B(\beta,
\alpha; 0) = \Pi_A(\alpha, \beta; 0) $, provided that the payoff
operators $A$ and $B$ fulfill $B = S\, A\, S.$

 Analogously, if we introduce the notation $\bar i = 1 - i$ for $i = 0, 1$ , i.e.
$\bar 0 = 1$ and $\bar 1 = 0$ and thereby the twist operator $ T |
i, j  \rangle  = | \bar j, \bar i \rangle, $ and the twisted
states, $ \left | \bar\beta, \bar\alpha \right> := T  \left |
\alpha, \beta \right> = \sum_{i, j} \xi_i(\alpha)\,
\chi_j(\beta)\, | \bar j, \bar i  \rangle , $ we find that in the
classical limit the game is $T$-symmetric, $\Pi_B(\bar\beta,
\bar\alpha; 0) = \Pi_A(\alpha, \beta; 0) $, provided that the
operators fulfill $ B = T\, A\, T $.

The symmetries can be elevated to the full quantum level if we
adopt the correlation factor in the form $
 J(\gamma) =
e^{i \gamma_1S/2}e^{i \gamma_2T/2},$ with real parameters
$\gamma_i \in [0, 2\pi)$ for $i = 1, 2.$

Since the individual strategies $\left | \alpha \right>_A$ and $
\left | \beta \right>_B$ are specified by $2 + 2 = 4$ parameters,
the correlation factor must have another $2$ parameters to cover
the full joint strategy space. The actual construction of the
correlation factor is far from unique, and our form is adopted
based on the convenience for the duality map..
In fact, one can readily confirm, using $[S, T] = ST - TS = 0$,
that under the game is $S$-symmetric $ \Pi_B(\beta, \alpha;
\gamma) = \Pi_A(\alpha, \beta; \gamma), \label{spayoffrel} $ even
in the presence of the correlation .

Similarly, the game is $T$-symmetric $ \Pi_B(\bar\beta,
\bar\alpha; \gamma) = \Pi_A(\alpha, \beta; \gamma)$ .

Since the correlation parameters in $\gamma$ are arbitrary, the
properties of games imply that a symmetric quantum game consists
of a ($\gamma$-parameter) family of games with the ($S$ or $T$)
symmetry exhibited for each $\gamma$.

It is interesting to observe that these two types of symmetric
games are actually related by unitary transformations. To see
this, let us introduce the operator $C_A$ which implements the
conversion for Alice's strategies,
 $ C_A \{i, j\} =
\{\bar{i}, j\}.  $

Note that $C_A$ satisfies $ C_A\, S\, C_A = T, \qquad C_A\, T\,
C_A = S$. Consider then the transformation of strategy by
unilateral conversion by Alice, $\{\alpha, \beta; \gamma\}$ to
$C_A \{\alpha, \beta; \gamma\}$.

On account of the relation  and the form of the correlation, we
find

$ C_A \{\alpha, \beta; \gamma\} = \{\bar\alpha, \beta;
\bar\gamma\}, $ with $\bar\gamma$ given by $(\bar\gamma_1,
\bar\gamma_2) = (\gamma_2, \gamma_1)$.

In addition, one may also consider the transformation on the
payoff operators,  $A$ to $\bar A = C_A\, A\,C_A, \quad B$ to
$\bar B = C_A\, B\, C_A$.

One then observes that, if the game is $S$-symmetric fulfilling
the game defined by the transformed operators becomes
$T$-symmetric, $\bar B = T\, \bar A\, T$.

Analogously, if the game is $T$-symmetric, then the transformed
operators define an $S$-symmetric game, $\bar B = S\, \bar A\, S.$

This shows that the conversion $C_A$ in  provides a one-to-one
correspondence, or {\it duality}, between an $S$-symmetric game
and a $T$-symmetric game. Some quantities in quantum game are
invariant under the duality map while other are not.   For
instance, the trace of the payoff, $ {\rm Tr}\,A = \sum_{i,j}
A_{ij}= A_{00} + A_{01} +A_{10} +A_{11}, $ remains invariant ${\rm
Tr}\,A \to {\rm Tr}\,\bar A = {\rm Tr}\,A$, whereas the alternate
trace defined by $ \!\!\!\!\! \tau (A)\sum_{i,j} (-)^{i+j}A_{ij}=
A_{00} - A_{01} -A_{10} +A_{11}, \label{eq:trickone} $ changes the
sign $\tau (A)  \to \tau (\bar A) = - \tau (A)$.

In formal terms, the two games given by $\{{\cal H}, A, B\}$ and
$\{{\cal H}, \bar A, \bar B\}$ are dual to each other in the sense
that the payoff under the strategy $ \{\alpha, \beta; \gamma\}$ in
one game is equivalent to the payoff under the dual strategy $C_A
\{\alpha, \beta; \gamma\} = \{\bar\alpha, \beta; \bar\gamma\}$ in
the other.

In particular, if the former game happens to be $S$-symmetric,
then the latter is $T$-symmetric, and vice versa. This allows us
to regard any two games as \lq identical\rq\ if their payoff
operators are related by the duality map.

Evidently,  the other conversion of the strategies by Bob $C_B
\{i, j\} = \{i, 1- j\}$ can also be used to provide a similar but
different duality.  Besides, their combination, $ C= C_A\otimes
C_B  $ implements the renaming of the strategies $0
\leftrightarrow 1$ for both of the players, and yields a duality
map which does not alter the type of symmetries of the game. These
duality maps $C_A$, $C_B$ and $C$ are used later to identify games
defined from different classical payoff matrices. We mention that
these dualities are actually a special case of the more general
\lq gauge symmetry\rq\ in quantum game theory, which is that the
two games defined by $\{{\cal H}, A, B\}$ and $\{{\cal H}, U A
U^\dagger, U B U^\dagger\}$ with some unitary operator $U$ are
dual to each other under the corresponding strategies $ \{\alpha,
\beta; \gamma\}$ and $U \{\alpha, \beta; \gamma\}$.

The matrix $U \{\alpha, \beta; \gamma\}$ and the corresponding
strategies $ \{\alpha, \beta; \gamma\}$ can be selected as
super-gauge and corresponding super-strategies.

Thus the identification of games can be extended to those which
are unitary equivalent.

In the case of SSQM it is the case of intertwined operators which
have almost the same spectrum.

\section{Conclusions}

The application an idea of the quantum game theory, which is now
popular in many branches of science and physics, for  cryptology
can give quite new possibilities.

Quantum games may lead to a deeper understanding of quantum
algorithms and quantum information processing and could even shed
light not only on the great difference between quantum and
classical physics, but say something novel for games in the
Nature.

It is fundamental to view unitary braiding operators describing
topological entanglements as universal quantum gates for quantum
computation.Entanglement is playing main role in quantum game
theory and via braiding is connected with supersymmetry in the
following sense:

a generalization of the category of super vector spaces in which
the grading is replaced by a braiding is the braided category and
in this sense one way to study topological entanglement and
quantum entanglement is to try making direct correspondences
between patterns of topological linking and entangled quantum
states.

The $2\otimes2$-dimensional quantum systems are of particular
interest and importance to the study of quantum games.

Some speculations with the connection of the supersymmetry and
quantum games gives possibility to obtain full symmetric games
which can be unitary equivalent like supersymmetric spectra
(isospectral supersymmetric Hamiltonians).

Important is the result that the identification of quantum games
can be extended to those which are unitary equivalent and so
obtain winning strategies for unitary equivalent quantum games.
This can be applied in the cryptology looking for the winning
strategies in codebreaking.

The entanglement of two the $2\otimes2$-dimensional quantum
systems can be important for the study symmetries of quantum
games.

   This work was supported by Grant T300100403 GA AV CR  .

\end{document}